\documentclass[%
 reprint,
 superscriptaddress,
 amsmath,amssymb,
 aps,
 prc,
floatfix,
]{revtex4-1}

\usepackage{graphicx}
\usepackage{dcolumn}
\usepackage{ulem}
\usepackage{bm}
\usepackage{lipsum}

\usepackage{color}

\usepackage{multirow}

\begin{document}

\preprint{APS/123-QED}

\title{Nuclear Transparency in Monte Carlo Neutrino Event Generators}

\author{Kajetan Niewczas}
 \email{kajetan.niewczas@uwr.edu.pl}
 \affiliation{Institute of Theoretical Physics, University of Wroc{\l}aw, Plac Maxa Borna 9, 50-204, Wroc{\l}aw, Poland}
 \affiliation{Department of Physics and Astronomy, Ghent University, Proeftuinstraat 86, B-9000 Gent, Belgium}
\author{Jan T. Sobczyk}
 \affiliation{Institute of Theoretical Physics, University of Wroc{\l}aw, Plac Maxa Borna 9, 50-204, Wroc{\l}aw, Poland}

\date{\today}

\begin{abstract}
Hadron cascade model is an essential part of Monte Carlo neutrino event generators that governs final state interactions of knocked-out nucleons and produced pions. It is shown that such model enriched with physically motivated modifications of nucleon-nucleon cross section and incorporation of nuclear  correlation effects is able to reproduce experimental nuclear transparency data. Uncertainty of nucleon final state interactions effects is estimated and applied to recent neutrino-nucleus cross section measurements that include an outgoing proton in the experimental signal. Conclusions are drawn on a perspective of identification of events that originate from two-body current mechanism.
\end{abstract}

\maketitle


\section{Introduction}
\label{sec:introduction}

The description of the transport of hadrons in nuclear matter is a challenge encountered in many areas of fundamental research such as astrophysics, phenomenology of heavy-ion collisions, as well as in a broad spectrum of nuclear physics applications.
The first notable attempt of modeling this process was a Monte Carlo (MC) approach based on the ideas of Serber~\cite{Serber:1947zza} and implemented by Metropolis et al.~\cite{Metropolis:1958wvo}.
This concept of cascading hadrons was later followed by many others and the model developed significantly~\cite{Bertini:1970zs, Cugnon:1980zz, Gudima:1983zz, Bohlen:2014buj}.
However, cascade models are based on theoretical assumptions which put limitations on their applicability \cite{Serber:1947zza, Yariv2007}.
This motivated development of two alternative solutions, that go beyond this simplified picture and  were successfully used in, e.g., heavy-ion physics.
The first one is based on \mbox{Boltzmann-Uehling-Uhlenbeck} (BUU) equations~\cite{Buss:2011mx}, formulated for the evolution of the \mbox{one-body} \mbox{phase-space} density under the influence of a mean field.
The second one, known under the name of Quantum Molecular Dynamics (QMD)~\cite{Aichelin:1991xy}, is formulated in terms of nucleon coordinates and momenta under the action of a \mbox{many-body} Hamiltonian. 
Both approaches are supplemented with a \mbox{two-body} collision term.
A detail comparison of 15 independent implementations of BUU and QMD models shows surprisingly large differences in their predictions~\cite{Zhang:2017esm}.

The transport problem  is also of great importance for investigation of elementary projectiles scattering off atomic nuclei that involve final-state interactions (FSI) of knocked-out nucleons and produced mesons.
It is particularly relevant in neutrino physics, where further progress in reducing systematic errors in long- and \mbox{short-baseline} oscillation experiments~\cite{Alvarez-Ruso:2017oui} requires a more extensive use of measurements of final state protons.
On one hand, it is needed in calorimetric techniques to reconstruct neutrino energies.
It is well established that more exclusive final state measurements provide a better estimation of the neutrino energy~\cite{Mosel:2013fxa, Furmanski:2016wqo}.
Also, the investigation of final state protons allows to learn about a size of multinucleon ejection contribution to the inclusive cross section which is important even if neutrino energy is reconstructed based on the observation of  the final state muon only.

Analyses of oscillation experiments require reliable theoretical predictions for the complexity of nuclear response to neutrino probes with broad energetic spectra.
It is done using event generators, such as NEUT, GENIE, NuWro, and GiBUU~\cite{Patrignani:2016xqp}.
The first three of them model FSI using the intranuclear cascade model.
An important test that  the FSI model should pass is the ability to reproduce nuclear transparency data from electron scattering studies~\cite{Dutta:2012ii}.
Nuclear transparency is defined as a  probability that a \mbox{knocked-out} nucleon is not subject to \mbox{re-interactions} inside the residual nucleus.
In the case of carbon target, used in MINERvA~\cite{Aliaga:2013uqz} and T2K~\cite{Abe:2011ks} experiments, typical transparency values are of the order of $65\%$ and in the vast majority of events, \mbox{knocked-out} nucleons interact at most once.

The goal of this paper is to present a procedure of checking if neutrino MC event generators reproduce nuclear transparency data.
The discussion is done using NuWro generator~\cite{Golan:2012wx} but can easily be repeated with other neutrino MCs.
Our main conclusion is that NuWro nucleon cascade model, after enrichment of its physical content, describes nuclear transparency very well.

By comparing NuWro results with the electron transparency data we also estimated the uncertainty of nucleon mean free path as calculated in the NuWro cascade model.
Our conclusion is that if it is scaled up and down by $\sim 30\%$ one gets an error bound that covers experimental uncertainties.
With this estimate, we investigated how important is an impact of the FSI uncertainty for the recent T2K proton measurements. This defines a bound of experimental sensitivity in attempts to draw conclusions about \mbox{two-body} current mechanism. 
Our conclusion is that nucleon FSI effects are controlled well enough to make a possibility  of investigating details of multinucleon ejection dynamics realistically.

The paper is organized as follows.
In section~\ref{sec:transparency} we describe how nuclear transparency is defined and measured in electron scattering experiments, following the procedures described in Refs.~\cite{Abbott:1997bc, Dutta:2003yt}.
In section \ref{sec:nuwro}, NuWro generator is presented with a focus on the description of the nucleon cascade model.
Section~\ref{sec:results} contains the details of transparency computations in NuWro.
The last two Secs.:~\ref{sec:discussion} and~\ref{sec:conclusions} 
contain a discussion of the results and our conclusions.

\section{Nuclear transparency}
\label{sec:transparency}

Modeling of scattering processes on nuclear targets strongly relies on a description of nucleon propagation within nuclear medium.
To estimate the magnitude of nucleon distortion, one can introduce a measure, called nuclear transparency, defined as the probability of a struck nucleon to escape nucleus without significant \mbox{re-interactions}.
Much attention has been brought to this subject following the hypothesis of color transparency (CT)~\cite{Dutta:2012ii}.
Such phenomenon should suppress the probability of \mbox{in-medium} \mbox{nucleon-nucleon} interaction at very high energies.
CT has been extensively studied in many experiments, using \mbox{quasi-free} $A$($e$,$e^\prime p$) scattering on various nuclei, so far without definite conclusions~\cite{Dutta:2012ii}.

The general idea behind the measurement of the nuclear transparency in \mbox{quasi-elastic} $A$($e$,$e^\prime p$) reactions is to confront an experimental yield of \mbox{knocked-out} protons with a theoretical prediction that does not include the distortion due to FSI.
In these experiments, where an electron ejects a proton $p$ out of a nucleus $A$, using measured values of energy $\omega=E-E'$ and momentum $\vec{q}=\vec{k}-\vec{k'}$ transfers ($E$ and $\vec{k}$ are the initial electron energy and momentum, primed values refer to the final electron), one defines the missing momentum and the missing energy as follows:
\begin{align}
\vec{p}_m &\equiv \vec{p}_p - \vec{q},\\
E_m &\equiv \omega - T_p - T_{A-1}.
\end{align}
Here $T_p$ and $T_{A-1} = |\vec{p}_m|^2/2M_{A-1}$ are the kinetic energies of the knocked-out proton and the residual nucleus, respectively.
The nuclear transparency, measured for a fixed \mbox{four-momentum} transfer $Q^2 \equiv |\vec{q}|^2 - \omega^2$, is defined as
\begin{equation}
\label{eq:transparency}
T(Q^2) = \frac{\int_V \mathrm{d}^3p_m\mathrm{d}E_m \ Y_\mathrm{exp}(E_m,\vec{p}_m)}{\int_V \mathrm{d}^3p_m\mathrm{d}E_m \ Y_\mathrm{PWIA}(E_m,\vec{p}_m)},
\end{equation}
where $Y_\mathrm{exp}$ and $Y_\mathrm{PWIA}$ are proton yields of the measurement and theoretical calculation, respectively.
The phase space $V$ is restricted to the \mbox{quasi-elastic} region by the conditions $E_m \lesssim 80$~MeV and $|\vec{p}_m| \lesssim 300$~MeV, which ensure a suppression of inelastic processes.
The theoretical prediction $Y_\mathrm{PWIA}$ is calculated under a hypothesis of the plain wave impulse approximation (PWIA), i.e., that the knocked-out nucleon does not undergo any \mbox{re-interactions}.
One should be aware that the aforementioned definition suffers from a model dependency as it relies on the accuracy of theoretical PWIA computations.

Over the years, the following experiments have reported nuclear transparency measurements:
\begin{itemize}
 \item {\it D.F. Geesaman, G. Garino et al.} at Bates Linear Accelerator Center~\cite{Geesaman:1989nt,Garino:1992ca},
 \item NE-18 at Stanford Linear Accelerator Center~\cite{Makins:1994mm,ONeill:1994znv},
 \item E91-013 in Hall C at Thomas Jefferson National Accelerator Facility~\cite{Abbott:1997bc,Dutta:2003yt},
 \item E94-139 in Hall C at Thomas Jefferson National Accelerator Facility~\cite{Garrow:2001di},
 \item E97-006 in Hall C at Thomas Jefferson National Accelerator Facility~\cite{Rohe:2005vc}.
\end{itemize}
The measurements were done in different kinematical setups, with outgoing protons momenta in the range from $\sim 0.5$ to $\sim 5.5~\mathrm{GeV/c}$, and for various nuclear targets, with the most widely used $^{12}$C and $^{56}$Fe.
The information about the kinematics of transparency measurements is summarized in Table~\ref{tab:kinematics}.

\begin{table}[t]
 \begin{tabular}{|c|c c c c c|}
  \hline
   \multirow{4}{*}{Ref.} & & Central & Central & Central & Central\\
   & Beam & electron & electron & proton & proton\\
   & energy & energy & angle & momentum & angle\\
   & (MeV) & (MeV) & (deg) & (MeV/c) & (deg)\\
   \hline \hline
   \multirow{2}{*}{\cite{Geesaman:1989nt,Garino:1992ca}} & \multirow{2}{*}{780} & \multirow{2}{*}{565} & \multirow{2}{*}{50.3} & \multirow{2}{*}{572.5} & 50.1, 58.2,\\
   & & & & & 67.9, 72.9\\
  \hline \hline
   \multirow{8}{*}{\cite{Makins:1994mm,ONeill:1994znv}} & \multirow{3}{*}{2015} & \multirow{3}{*}{1390} & \multirow{3}{*}{35.5} & \multirow{3}{*}{1200} & 43.4, 46.2,\\
   & & & & & 49.0, 51.8,\\
   & & & & & 54.6\\
   \cline{2-6}
   & \multirow{2}{*}{3188} & \multirow{2}{*}{1470} & \multirow{2}{*}{47.7} & \multirow{2}{*}{2450} & 27.7, 30.5,\\
   & & & & & 33.3\\
   \cline{2-6}
   & \multirow{1}{*}{4212} & \multirow{1}{*}{1470} & \multirow{1}{*}{53.4} & \multirow{1}{*}{3540} & 20.9, 22.6\\
   \cline{2-6}
   & \multirow{2}{*}{5120} & \multirow{2}{*}{1470} & \multirow{2}{*}{56.6} & \multirow{2}{*}{4490} & 15.9, 16.7,\\
   & & & & & 17.3\\
  \hline \hline
   \multirow{20}{*}{\cite{Abbott:1997bc,Dutta:1999iv,Dutta:2003yt}}
   & \multirow{6}{*}{2445} & \multirow{6}{*}{2075} & \multirow{6}{*}{20.5} & \multirow{6}{*}{882.8} & 35.4, 39.4,\\
   & & & & & 43.4, 47.4,\\
   & & & & & 51.4, 55.4,\\
   & & & & & 59.4, 63.4,\\
   & & & & & 67.4, 71.4,\\
   & & & & & 75.4\\
   \cline{2-6}
   & \multirow{3}{*}{3245} & \multirow{3}{*}{2255} & \multirow{3}{*}{28.6} & \multirow{3}{*}{1661.7} & 32.6, 36.6,\\
   & & & & & 40.6, 44.6,\\
   & & & & & 48.6, 52.6\\
   \cline{2-6}
   & \multirow{4}{*}{2445} & \multirow{4}{*}{1755} & \multirow{4}{*}{32.0} & \multirow{4}{*}{1343} & 31.5, 35.5,\\
   & & & & & 39.5, 43.5,\\
   & & & & & 47.5, 51.5,\\
   & & & & & 55.5\\
   \cline{2-6}
   & \multirow{2}{*}{3245} & \multirow{2}{*}{1400} & \multirow{2}{*}{50.0} & \multirow{2}{*}{2572.5} & 25.5, 28.0,\\
   & & & & & 30.5\\
   \cline{2-6}
   & \multirow{3}{*}{845} & \multirow{3}{*}{475} & \multirow{3}{*}{78.5} & \multirow{3}{*}{882.8} & 27.8, 31.8,\\
   & & & & & 35.8, 39.8,\\
   & & & & & 43.8, 47.8\\
   \cline{2-6}
   & \multirow{2}{*}{1645} & \multirow{2}{*}{675} & \multirow{2}{*}{80.0} & \multirow{2}{*}{1661.7} & 22.8, 26.8,\\
   & & & & & 30.8, 34.8\\
  \hline \hline
   \multirow{5}{*}{\cite{Garrow:2001di,McKee:2003va}} & \multirow{3}{*}{3059} & \multirow{3}{*}{1300} & \multirow{3}{*}{54.0} & \multirow{3}{*}{2520} & 19.8, 22.3,\\
   & & & & & 24.8, 27.3,\\
   & & & & & 29.8\\
   \cline{2-6}
   & \multirow{1}{*}{4463} & \multirow{1}{*}{1200} & \multirow{1}{*}{64.6} & \multirow{1}{*}{4090} & 15.3\\
   \cline{2-6}
   & \multirow{1}{*}{5560} & \multirow{1}{*}{1270} & \multirow{1}{*}{64.6} & \multirow{1}{*}{5150} & 12.8\\
  \hline \hline
   \multirow{5}{*}{\cite{Rohe:2005vc}} & \multirow{1}{*}{3298} & \multirow{1}{*}{2950} & \multirow{1}{*}{14.4} & \multirow{1}{*}{850} & 60.3\\
   \cline{2-6}
   & \multirow{1}{*}{3298} & \multirow{1}{*}{2750} & \multirow{1}{*}{17.0} & \multirow{1}{*}{1000} & 56.2\\
   \cline{2-6}
   & \multirow{1}{*}{3123} & \multirow{1}{*}{2500} & \multirow{1}{*}{22.2} & \multirow{1}{*}{1250} & 49.7\\
   \cline{2-6}
   & \multirow{1}{*}{3298} & \multirow{1}{*}{2400} & \multirow{1}{*}{25.4} & \multirow{1}{*}{1500} & 44.6\\
   \cline{2-6}
   & \multirow{1}{*}{3298} & \multirow{1}{*}{2280} & \multirow{1}{*}{29.0} & \multirow{1}{*}{1700} & 40.7\\
  \hline
 \end{tabular}
 \caption{Kinematical setups of $A$($e$,$e^\prime p$) experiments that reported nuclear transparency measurements.}
 \label{tab:kinematics}
\end{table}

The PWIA models used by experimental groups describe the proton target in the independent particle shell models (IPSM).
The IPSM based calculations are known to overestimate \mbox{single-particle} strength in exclusive reactions~\cite{Udias:1993xy}.
This discrepancy is attributed to the shells that are not fully occupied due to \mbox{nucleon-nucleon} correlations that cannot be fully accounted for in \mbox{mean-field} approaches.
\mbox{NE-18} at SLAC was the first experiment that introduced correlation factors $c_A$ in the definition of transparency to correct for the depletion of \mbox{single-particle} strength outside of the phase space $V$:
\begin{equation}
Y_\mathrm{PWIA}(E_m,\vec{p}_m) = c_A \ Y_\mathrm{IPSM}(E_m,\vec{p}_m)
\end{equation}
with values $c_A = 0.90$, $0.82$ for $^{12}$C and $^{56}$Fe, respectively.
They are larger than typically used spectroscopic factors, as they come up from the integration over a specific phase space $V$~\cite{Rohe:2005vc}.
In this paper we compare our results to transparency results as they were published by experimental groups.
Our treatment of correlation factors agrees with that from Ref.~\cite{Benhar:2006hh}.
 
It has to be emphasised that the introduction of correlation factors is a subject of an ongoing debate.
Some authors argue that because the experiments were conducted in the transverse kinematics, which is less sensitive to the high-value tail of the nucleon momentum distribution, the use of correlation factors is not justified~\cite{Frankfurt:2000ty}.
Theoretical arguments suggest that perhaps soft $Q^2$ dependent correlation factors would be more appropriate~\cite{Lapikas:1999ss}, but many recent papers on the nuclear transparency simply ignore them~\cite{Dutta:2012ii}.
CLAS Collaboration measured nuclear transparency of protons from \mbox{short-range} correlated (SRC) pairs~\cite{Hen:2012yva} and arrived at the conclusion that transparency ratios Al/C, Fe/C and Pb/C are consistent with the absence of the correlation factors in the definition Eq.~\ref{eq:transparency}.
Similar conclusion is also supported by theoretical computations based on the Glauber theory~\cite{Lava:2004zi, Martinez:2005xe, Cosyn:2013qe} and the relativistic optical potential~\cite{Lava:2004zi, Martinez:2005xe, Kelly:2005is}.

\section{NuWro}
\label{sec:nuwro}

\begin{table*}
 \begin{tabular}{|c|c c c c c c c c|}
  \hline
   \multirow{4}{*}{Ref.} & & Electron & Electron & Proton & Proton & Common & Missing & Missing\\
   & Targets & energy & angle & momentum & angle & plane & energy & momentum\\
   & used & acceptance & acceptance & acceptance & acceptance & acceptance & cut & cut\\
   & & (\%) & (deg) & (\%) & (deg) & (deg) & (MeV) & (MeV/c)\\
   \hline \hline
   \multirow{1}{*}{\cite{Geesaman:1989nt,Garino:1992ca}} & \multirow{1}{*}{C, Fe} & \multirow{1}{*}{3.5} & \multirow{1}{*}{1.4} & \multirow{1}{*}{25} & 1.1 & 10.8 & 80 & -\\
  \hline \hline
   \multirow{1}{*}{\cite{Makins:1994mm,ONeill:1994znv}} & \multirow{1}{*}{C, Fe} & \multirow{1}{*}{5} & \multirow{1}{*}{0.9} & \multirow{1}{*}{5} & 0.9 & 4.6 & 100 & 250\\
  \hline \hline
   \multirow{1}{*}{\cite{Abbott:1997bc,Dutta:1999iv,Dutta:2003yt}} & \multirow{1}{*}{C, Fe} & \multirow{1}{*}{10} & \multirow{1}{*}{2.4} & \multirow{1}{*}{20} & 3.4 & 4.6 & 80 & 300\\
  \hline \hline
   \multirow{1}{*}{\cite{Garrow:2001di,McKee:2003va}} & \multirow{1}{*}{C, Fe} & \multirow{1}{*}{15} & \multirow{1}{*}{3.4} & \multirow{1}{*}{8} & 2.4 & 9.4 & 80 & 300\\
  \hline \hline
   \multirow{1}{*}{\cite{Rohe:2005vc}} & \multirow{1}{*}{C} & \multirow{1}{*}{9.6} & \multirow{1}{*}{2.4} & \multirow{1}{*}{15} & 3.4 & 7 & 80 & 300\\
  \hline
 \end{tabular}
 \caption{Table of cuts used by experimental groups and introduced in our simulations. }
 \label{tab:cuts}
\end{table*}

NuWro~\cite{Golan:2012wx} is a neutrino Monte Carlo generator that has been developed at University of Wroc{\l}aw since 2005.
It covers neutrino energy range from $\sim~100$~MeV to $\sim~100$~GeV.
For \mbox{neutrino-nucleon} scattering NuWro uses three interaction "modes": CCQE (or elastic for neutral current reaction); RES, which covers a region of invariant hadronic mass $W<1.6$~GeV; and DIS (jargon in the neutrino MC community for shallow and deep inelastic scattering), in which the inelastic processes have $W>1.6$~GeV.
In the case of \mbox{neutrino-nucleus} scattering it is assumed that interactions occur on bound and moving nucleons (impulse approximation).
A variety of options to describe such nucleons are available starting from global and local Fermi gas (LFG) models up to the hole spectral function (SF)~\cite{Benhar:1994hw} with the lepton affecting FSI effects included~\cite{Ankowski:2014yfa}, and density and momentum dependent potential~\cite{Juszczak:2005wk}.
The description of scattering off nuclear targets is completed with interactions mediated by meson exchange currents (MEC) and with the coherent pion production (COH).

For the purpose of this study the most important NuWro ingredient is the intranuclear cascade model described in the subsection~\ref{sec:sub:cascade}.

\subsection{NuWro cascade model}
\label{sec:sub:cascade}

The model describes \mbox{in-medium} propagation of pions and nucleons.
The scheme is taken from the seminal papers by Metropolis et al.~\cite{Metropolis:1958wvo, Metropolis:1958sb} but relevant physics ingredients are new.
The MC sampling is based on the standard formula that expresses probability for a particle to propagate over a distance $\Delta x$ with no re-interaction:
\begin{equation}
    P(\Delta x) = \exp(-\Delta x/\lambda),
    \label{eq:prob_cascade}
\end{equation}
where $\lambda = (\rho \sigma)^{-1}$ is the mean free path calculated locally, expressed in terms of nuclear density $\rho$ and an effective interaction cross section $\sigma$.
In actual computations we distinguish proton or neutron densities and proton-proton or proton-neutron cross sections.
A step of $\Delta x = 0.2 \ \mathrm{fm}$ was checked to be sufficient to grasp the structure of a nuclear density profile.

The performance of the NuWro pion cascade model was benchmarked on numerous neutrino-nucleus pion production cross section measurements showing in general a good agreement with the data, see e.g. Ref.~\cite{McGivern:2016bwh}.
In this study, we focus on the nucleon cascade model.

The computations in this paper were done with the NuWro version 19.02~\cite{NuWro} that contains several improvements, with respect to NuWro 18.02.
This version uses a custom fit to the experimental free \mbox{nucleon-nucleon} cross sections, both elastic and inelastic, that aimed to improve the agreement with the current PDG dataset~\cite{Patrignani:2016xqp}.
The fraction of \mbox{single-pion} production within inelastic interactions was adjusted to follow the fits of Ref.~\cite{Bystricky:1987yq}.
Moreover, the \mbox{center-of-momentum} (COM) frame angular distributions for the elastic scattering were updated using the parametrization of Ref.~\cite{Cugnon:1996kh}.

The in-medium modification of the elastic cross sections is modeled using the results of Pandharipande and Pieper study~\cite{Pandharipande:1992zz}, where
the two main effects come from Pauli blocking and \mbox{in-medium} nucleon effective mass.
The Pauli blocking is included on the \mbox{event-by-event} basis, a straightforward way in MC simulations.
We checked that NuWro cascade performance reproduces the results from Ref.~\cite{Pandharipande:1992zz} with a sufficient accuracy.
For the inelastic \mbox{nucleon-nucleon} scattering we adopt a phenomenological \mbox{in-medium} cross section ($\sigma^\ast_{\mathrm{NN}}$) parametrization~\cite{Klakow:1993dj}:
\begin{equation}
    \sigma^\ast_{\mathrm{NN}} = (1 - \eta \frac{\rho}{\rho_0}) \sigma^{\mathrm{free}}_{\mathrm{NN}},
\end{equation}
where $\eta = 0.2$, and $\rho$, $\rho_0$ are local and saturation nuclear densities, respectively.

Following the experiences of Refs.~\cite{Pandharipande:1992zz,Benhar:2006hh,Cosyn:2013qe}, we included effects coming from \mbox{nucleon-nucleon} \mbox{short-range} correlations. 
In general, the density that enters the mean free path in Eq.~\ref{eq:prob_cascade} is assumed to be the one of nuclear matter at point $\vec{r}_2$, as experienced by a propagating nucleon known to be in a position $\vec{r}_1$.
It can be expressed in terms of one- ($\rho^{[1]}_\mathrm{A}$) and \mbox{two-body} ($\rho^{[2]}_\mathrm{A}$) densities as
\begin{equation}
    \rho^{[1]}_{\mathrm{eff}}(\vec{r}_2|\vec{r}_1) = \frac{\rho^{[2]}_\mathrm{A}(\vec{r}_1,\vec{r}_2)}{\rho^{[1]}_\mathrm{A}(\vec{r}_1)},
    \label{eq:effdensity}
\end{equation}
normalized to the number of remaining nucleons $\int
\mathrm{d}^3\vec{r}_2 \
\rho^{[1]}_{\mathrm{eff}}(\vec{r}_2|\vec{r}_1) = A-1$.
We introduce correlation effects through a following substitution
\begin{equation}
    \begin{split}
    \rho^{[1]}_{\mathrm{eff,IPSM}}(\vec{r}_2|\vec{r}_1) & = \rho^{[1]}_\mathrm{A-1}(\vec{r}_2) \\
    & \to \rho^{[1]}_\mathrm{A-1}(\vec{r}_2) g(|\vec{r}_{21}|) N(|\vec{r}_2|),
    \end{split}
    \label{eq:effdensity_sub}
\end{equation}
where $g(|\vec{r}_{21}|)$ is the \mbox{nucleus-dependent} pair distribution function~\cite{Pandharipande:1992zz} and $N(|\vec{r}_2|)$ is introduced to keep the global normalization condition.

For the choice of $g(|\vec{r}_{21}|)$, we rely on distributions of \mbox{nucleon-nucleon} distances obtained in ab initio computations for light nuclei, including carbon~\cite{ANLdensity,Carlson:2014vla}.
For heavier nuclei, including iron, we approximate $g(|\vec{r}_{21}|)$ by the \mbox{ab initio-calculated} infinite nuclear matter distributions $g_{\mathrm{inf}}(\rho_{\mathrm{avg}},|\vec{r}_{21}|)$ of Ref.~\cite{Pandharipande:1992zz}, evaluated at average nuclear density.
In our computations we include effects coming from different shapes of $g(|\vec{r}_{21}|)$ for nucleon pairs of the distinct isospin configurations, and following the scheme summarized in~Eq.~\ref{eq:effdensity_sub} we define effective densities.

The discussion of the influence of the aforementioned NuWro cascade model modifications on the results of this paper can be found in Sec.~\ref{sec:sub:ingredients}.

\subsection{NuWro as a tool in transparency studies}
\label{sec:sub:trans}

Using MC event generators one can define the "MC transparency" as a fraction of events with no nucleon \mbox{re-interactions} at all.
However, experimentally one cannot distinguish these events from those with "soft" FSI.
Because of that, to make a comparison reliable, we go through all the steps of the experimental procedures to extract the theoretical counterpart of the measured transparency.

NuWro keeps the information about particles before and after FSI.
This is exactly what is needed in the computation of nuclear transparency.
Particles after FSI correspond to those that are detected  in experiments.
Particles before FSI correspond to theoretical computations in PWIA.

NuWro does not yet have a complete electron scattering module, hence in this study we use neutral current (NC) $\nu_e$ interactions on bound proton targets.
In doing so, we collect samples of NC events with exactly the same (electron mass is negligible) kinematics as in the transparency electron scattering experiments.
In both electron and neutrino cases, a radial distribution of interaction points inside the nucleus is the same and given by the nucleus density profile.

The main challenge is to reproduce experimental situations with complete information on the kinematics and applied cuts.
For every kinematical setup we ran a simulation with the neutrino beam energy equal to $E_e$.
Then, the energy $E_{e^\prime}$ and the \mbox{in-plane} angle $\theta_{e^\prime}$ for the  outgoing  electron or neutrino were fixed around the  central value of the spectrometer.
Analogically, the momentum $p_p$ and the \mbox{in-plane} angle $\theta_{p}$ for the \mbox{knocked-out} proton were fixed.
As in all the experiments the electron and proton spectrometers are set \mbox{in-plane}, the \mbox{out-of-plane} angles were fixed to the same value $\phi_{e^\prime} = \phi_p$.
The exclusive cross section formula is symmetric with respect to the rotation of the system, hence only the relative \mbox{out-of-plane} angle between the electron and proton plays a role, here set to $\phi_{e^\prime p} = 0$.
All of the variables $E_{e^\prime}, \theta_{e^\prime}, p_p, \theta_{p}, \phi_{e^\prime p}$ were fixed with the accuracy provided by the spectrometers' energy or angular acceptance, namely $\Delta E_{e^\prime}, \Delta \theta_{e^\prime}, \Delta p_p, \Delta \theta_{p}, \Delta \phi_{e^\prime p} = \Delta \phi_{e^\prime} + \Delta \phi_p$.
On the top of those cuts, additional conditions were imposed using the information about the $E_m, |\vec{p}_m|$.
The beam energies and central spectrometers values for every setup can be found in Tab.~\ref{tab:kinematics}, while the acceptances and the cuts on missing variables are put into Tab.~\ref{tab:cuts}.

\begin{figure}
  \centering
  \includegraphics[width=0.5\textwidth]{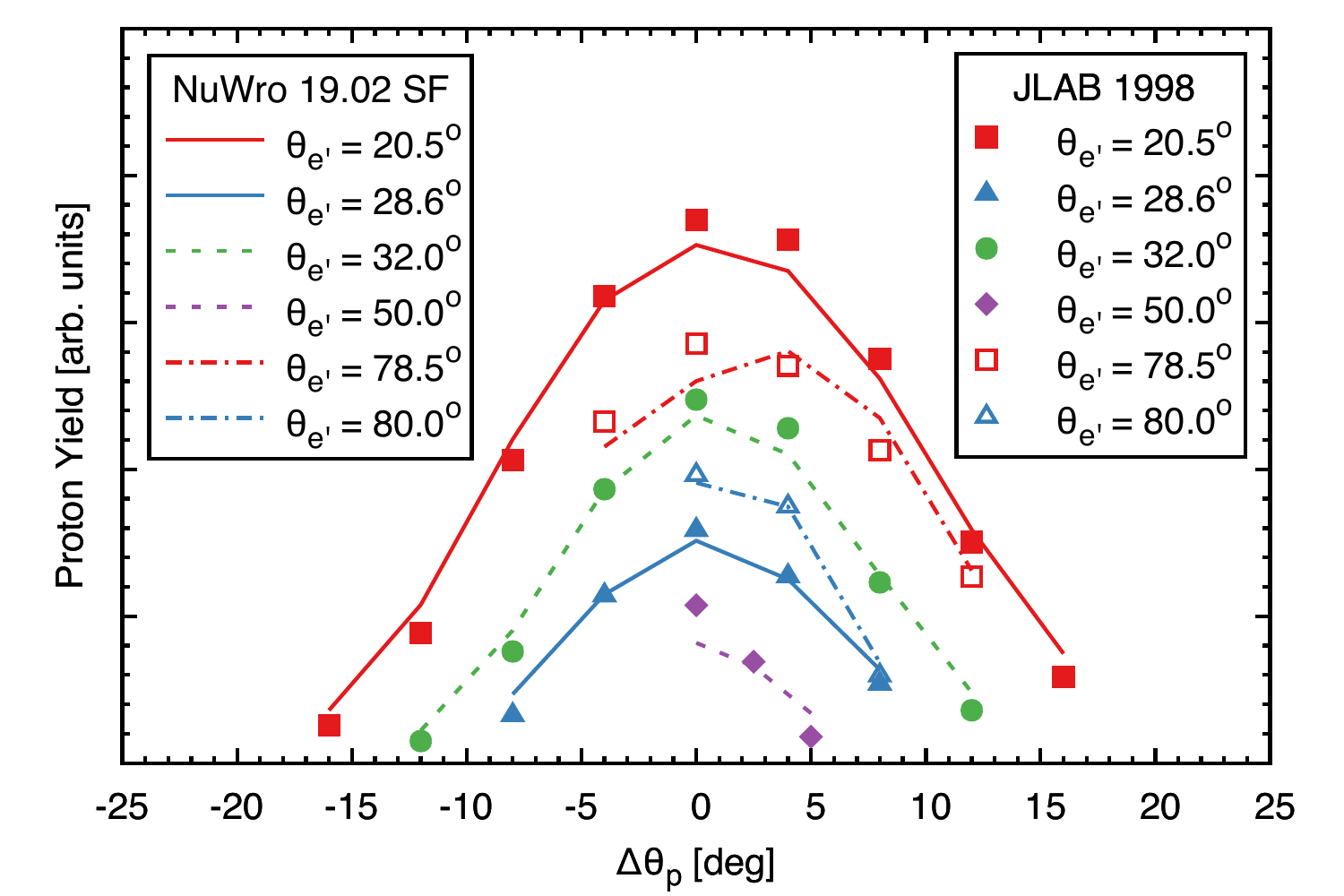}
  \caption{Angular distributions of the proton yield as measured in the E91-013 experiment at JLAB. Points represent data from Ref.~\cite{Dutta:2003yt} with the convention proposed in Ref.~\cite{Dutta:1999iv}. Lines are results computed with SF in NuWro. $\Delta \theta_p = 0$ corresponds to the free proton target case. Both distributions are normalized to the same area.}
  \label{fig:angular_dist}
\end{figure}

To establish a proper framework for comparing nuclear transparency results with experiment, we tested different ways of modeling the initial nuclear state in NuWro.
The SF- and \mbox{LFG-based} simulations were compared with exclusive properties of the \mbox{knocked-out} protons that were reported by the E91-013 experiment at JLAB.
As can be seen in Fig.~\ref{fig:angular_dist}, SF in NuWro is able to accurately reproduce a measured shape of the angular distributions of \mbox{knocked-out} protons.
The angular dependence of transparency reproduces a general flat shape that can be seen in Fig.~2. of Ref.~\cite{Abbott:1997bc} with sufficient precision. 
However, the angular distributions of the measured yield of protons for the \mbox{LFG-based} simulations is peaked too strongly around the central value, that leads to the overestimation of the proton transparency.
Due to its simplicity, LFG model fails to properly predict the exclusive kinematics, what is a prerequisite in reliable nuclear transparency studies.

We conclude that only NuWro simulations that uses SF as the model for the initial nuclear state can give reliable results in comparison with exclusive electron scattering experiments.
Unfortunately, such conclusion imposes a limitation on nuclear targets that can be simulated, as the hole spectral functions are available only for a limited number of nuclei making an estimation of the \mbox{A-dependence} of nuclear transparency impossible in NuWro.
The only targets that can be compared with the transparency measurements are $^{12}$C and $^{56}$Fe.

A similar study was done in the past using the Giessen BUU transport model~\cite{Leitner:2009ke, Lehr:phd}.
The experimental data from three JLab and SLAC experiments~\cite{Garrow:2001di, Abbott:1997bc, ONeill:1994znv} were analyzed using detailed information about angular acceptance of spectrometers. 
Interesting ingredients of the BUU discussion are: investigation of impact of restricted angular acceptance on final results, a study of transparency dependence on atomic mass $T(A)\sim A^{\alpha}$, an estimation of theoretical uncertainty due to not precisely known correlation factors $c_A$ Ref.~\cite{Frankfurt:2000ty}.

The final BUU results are similar to the ones presented in this paper as far as large proton momentum transparency saturation values are concerned.
However, there is a visible difference at lowest momentum ($Q^2$) experimental point: NuWro transparency continues to rise while the BUU transparency drops down.

\section{Results}
\label{sec:results}

\begin{figure}[t]
  \includegraphics[width=0.5\textwidth]{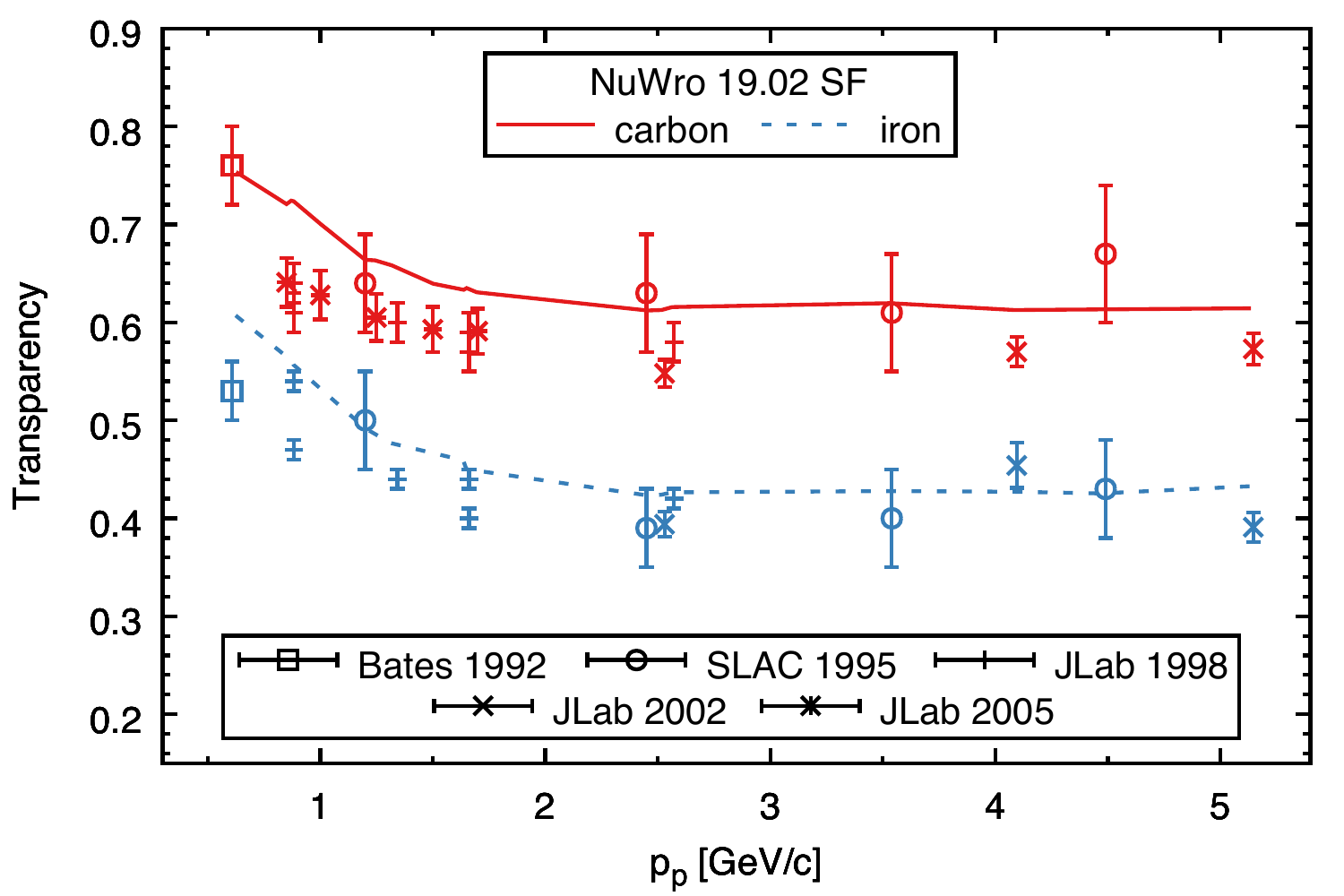}
  \caption{Nuclear transparency as a function of proton momentum. Lines represent results obtained with NuWro 19.02 using SF for carbon and iron targets. Experimental points come from  experimental papers mentioned in the text.}
  \label{fig:trans_final}
\end{figure}

In Fig.~\ref{fig:trans_final} the transparency results for carbon and iron are shown together with data points collected from several experiments.
In experimental papers, transparency is discussed as a function of $Q^2$ but this variable can be translated into an average proton momentum.
The transparency curve has a characteristic shape reproduced in all theoretical computations: a saturation at larger values of proton momentum and a decline in the region of $\sim 1$~GeV/c.
Saturation can be explained by a roughly constant value of the total free \mbox{nucleon-nucleon} cross sections for larger values of the incident nucleon momentum.
A region of transparency decline comes from a complicated interplay of various nuclear effects and is the most difficult to model.

NuWro simulations for carbon reproduce the transparency data quite well.
For application in neutrino physics, the most important region is that of low nucleon momentum, starting from $\sim 500$~MeV/c, which is a detection threshold in experiments like T2K and MINERvA.
We can see that the value of the first available experimental point, from Ref.~\cite{Garino:1992ca}, is reproduced well but then the decline of NuWro transparency is not steep enough.
Predictions from our model are slightly above the data in the saturation region.
For the iron target, the same shape of the transparency curve can be seen.
Small differences, including data overshooting at low momenta, can be attributed to \mbox{nucleon-nucleon} correlation effects being introduced in more approximate way with respect to carbon, see the discussion in \ref{sec:sub:cascade}.
In general, the agreement with the data points is satisfactory.

\subsection{Model uncertainties}

\begin{figure}[t]
  \includegraphics[width=0.5\textwidth]{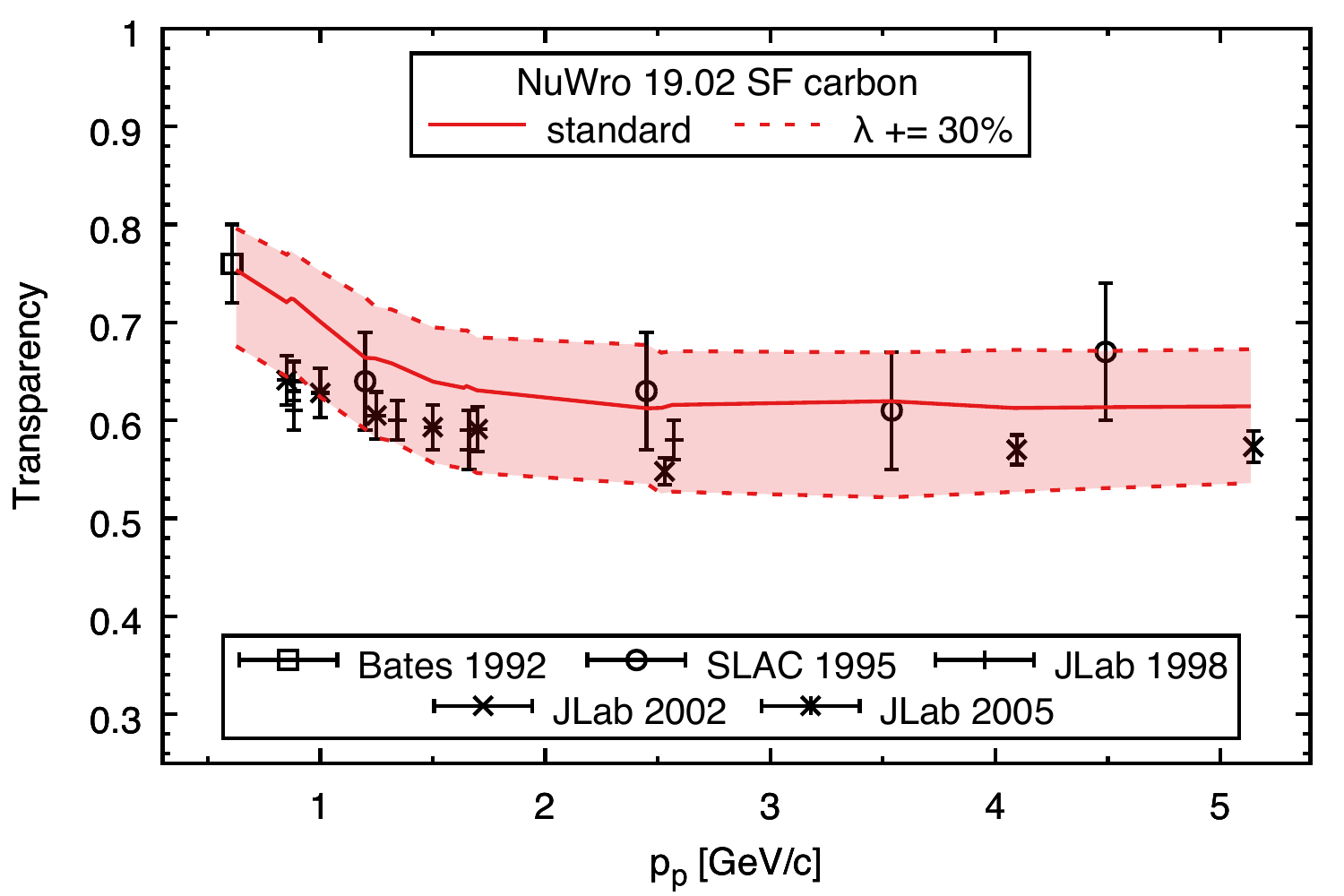}
  \caption{Nuclear transparency as a function of proton momentum. Lines represent results obtained with NuWro 19.02 using SF for carbon target. Dashed lines are results computed after scaling mean free paths by $\pm 30\%$. Experimental points come from the  papers mentioned in the text.}
  \label{fig:trans_final2}
\end{figure}

As discussed in the Introduction, nucleon FSI effects contribute to the background in all attempts to measure multinucleon ejection contribution to the inclusive cross section.
Thus, it is not enough to have good qualitative agreement with the transparency data but also it is important to estimate uncertainty inherent in the nucleon FSI model.
Our approach is to assess an uncertainty of the nucleon mean free path as calculated by NuWro.
We tried to define a 1$\sigma$ error bound by demanding that 2/3 of experimental points together with experimental errors are entirely inside the bound.
To achieve that we multiply mean free paths calculated within NuWro by a constant overall scaling factor.
The result is shown in Fig.~\ref{fig:trans_final2}.
The upper and lower dashed curves were obtained by scaling up and down central mean free paths by $30\%$.
A discussion of possible sources of uncertainty in NuWro FSI model is presented in Sec.~\ref{sec:sub:ingredients}.

\subsection{Monte Carlo transparency}
\label{sec:sub:mc_trans}

\begin{figure}[t]
  \includegraphics[width=0.5\textwidth]{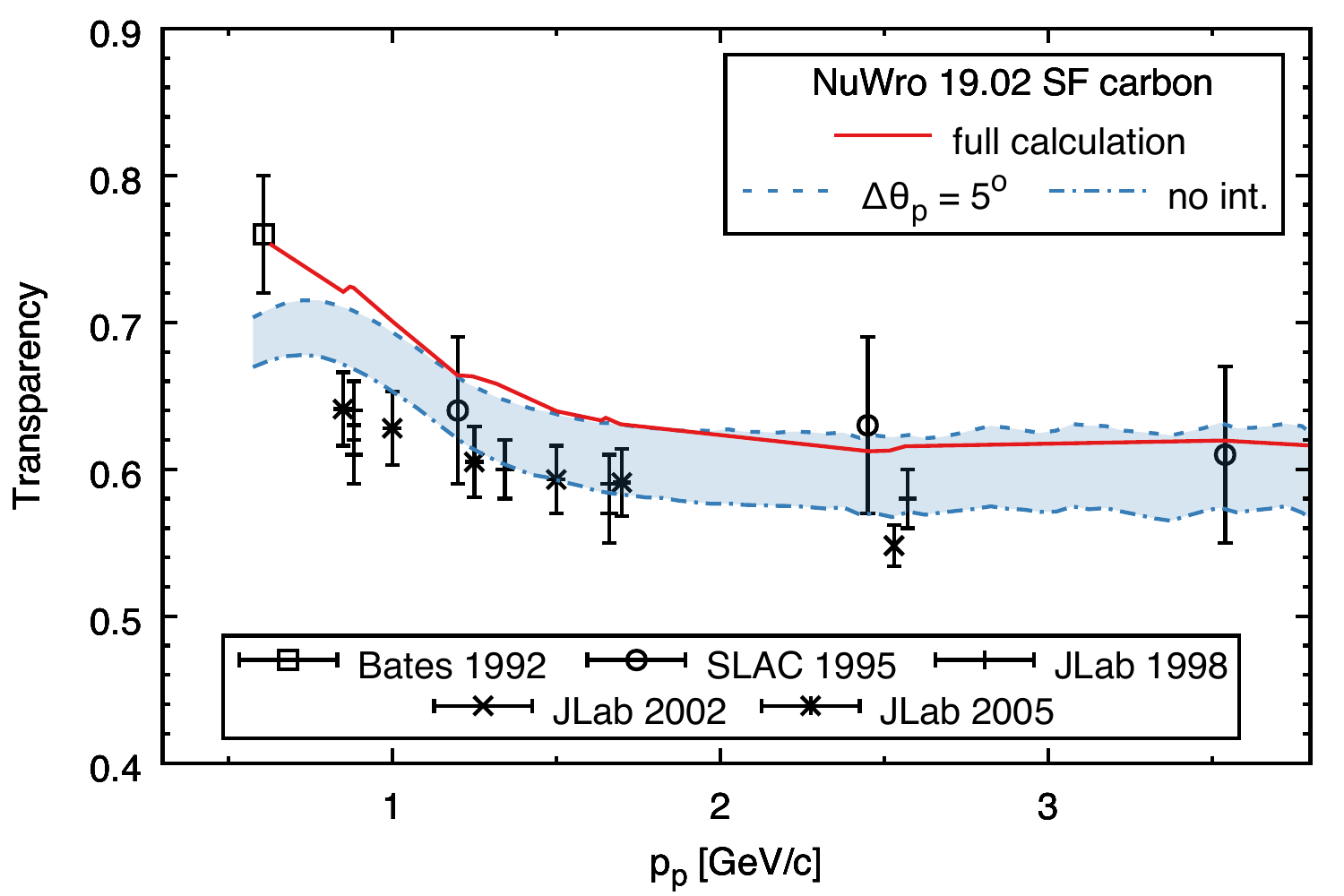}
  \caption{Nuclear transparency, calculated with the full model and with the approximation discussed in Sec.~\ref{sec:sub:mc_trans}, as a function of proton momentum.}
  \label{fig:trans_mcdef}
\end{figure}

In the MC approach, as mentioned in \ref{sec:sub:trans}, a natural way of studying the nuclear transparency is to follow individual cascaded protons and check whether they interact at all.
However as discussed earlier, such definition might not catch particular aspects of the situation that are important from the experimental perspective and is expected to underestimate the final result.
A refinement of the naive MC transparency definition is to take into account a finite angular acceptance of spectrometers, and therefore, allow protons to softly interact without a significant direction change, e.g., $\Delta \theta_p = 5^\circ$.
The value of $5^\circ$ approximately coincides with an angle that expands a solid angle in experimental acceptances, see Tab.~\ref{tab:cuts}.

In Fig.~\ref{fig:trans_mcdef}, the results for carbon, using different transparency definitions, are shown.
One can see that while the "no interactions" definition is too strict, the softer definition "$\Delta \theta_p = 5^\circ$" works quite well, especially in the saturation region.
However, it is unable to reproduce the first experimental point at $p_p \simeq 625 \ \mathrm{GeV/c}$.
Knowing this behavior, the definition "$\Delta \theta_p = 5^\circ$" can be used for less exhausting cascade checks.

\subsection{Cascade model ingredients}
\label{sec:sub:ingredients}

\begin{figure}[t]
  \includegraphics[width=0.5\textwidth]{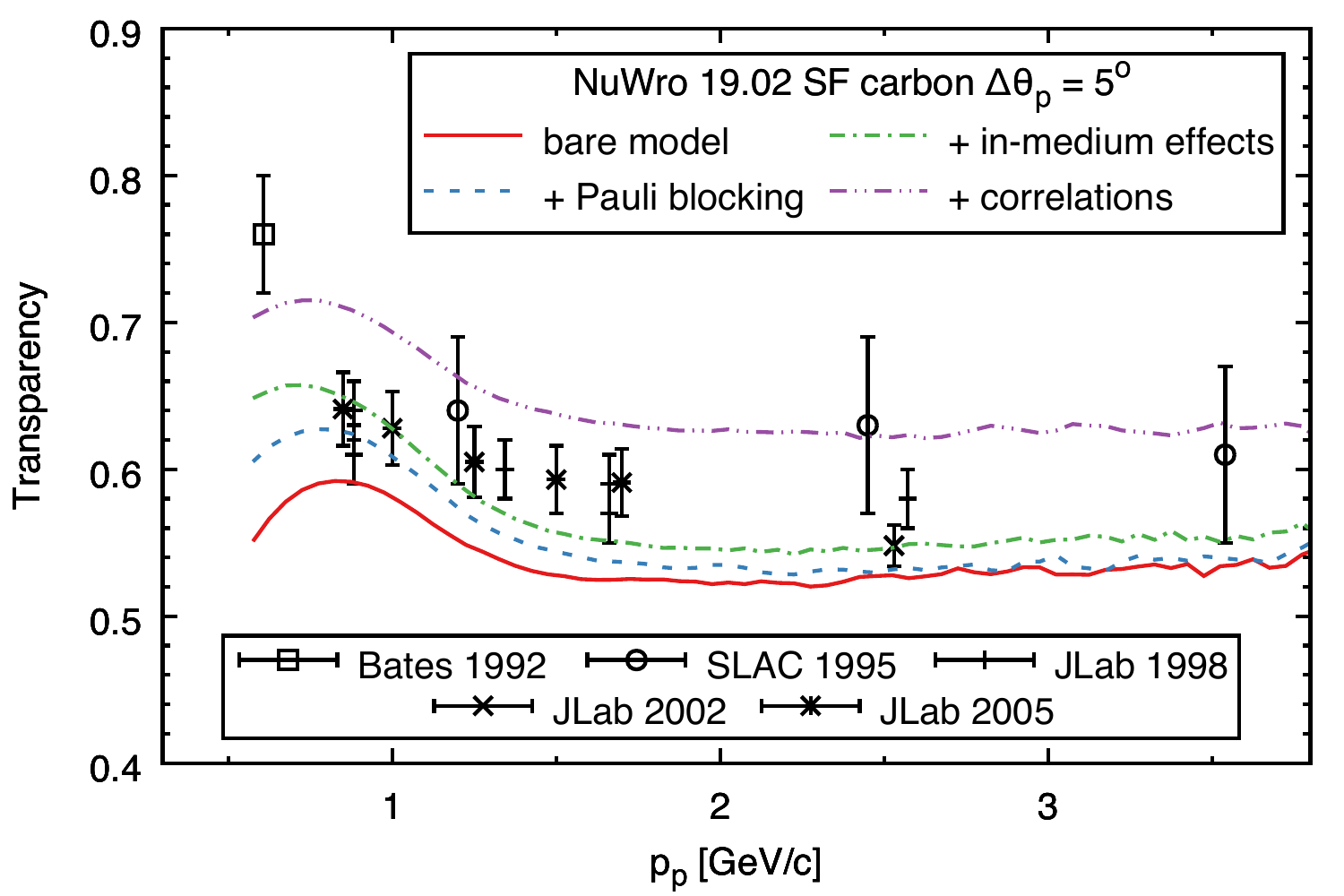}
  \caption{Nuclear transparency as a function of proton momentum obtained with different ingredients of the theoretical model, see explanations in the text.}
  \label{fig:trans_ingred}
\end{figure}

To understand sources of uncertainties in our model, we present an impact of its various ingredients on the predicted transparency.
In Fig.~\ref{fig:trans_ingred}, we show results obtained with:

\begin{itemize}
    \item a bare cascade model with free \mbox{nucleon-nucleon} cross sections, projectile binding energy and target nucleon Fermi motion effects,
    \item a model that on top of the bare model includes Pauli blocking (labeled "+ Pauli blocking"),
    \item a model that additionally includes \mbox{in-medium} \mbox{nucleon-nucleon} cross section effects (labeled "+ \mbox{in-medium} effects"),
    \item the full model that includes \mbox{nucleon-nucleon} correlation effects (labeled "+ correlations").
\end{itemize}

We can see that in the the region of proton momenta below 1~GeV/c all the theoretical ingredients of the model are relevant while for larger values of the momenta correlation effects play the most important role.

The basic observation about the bare model is that it underpredicts the experimentally measured transparency by a large amount.
The proton momentum dependence of the corresponding curve reflects the momentum dependence of free proton-proton or proton-neutron cross sections.
The effect of Pauli blocking is significant for lower momenta and slowly disappears with increasing proton momentum.
Although it might not seem to be intuitive that the impact of Pauli blocking extends up to $p_p \simeq 2.5 \ \mathrm{GeV/c}$, for larger elastic scattering energies, the COM angular distributions get more and more forward or backward peaked what leads to kinematics that are prone to be Pauli blocked.
As emphasised in Sec.~\ref{sec:sub:cascade}, the \mbox{in-medium} \mbox{nucleon-nucleon} cross section modifications are modeled differently for elastic and inelastic interactions.
This is reflected in nuclear transparency, where the modification of elastic cross sections has a stronger impact with lowering proton momentum, while the inelastic part has a constant behaviour.
The effect of the nuclear correlations strongly depends on average mean free paths in a given energy region.
The free \mbox{nucleon-nucleon} cross section is higher in the saturation region, and therefore, the mean free paths are lower and the effect of correlations is more pronounced.

In general, all of the more sophisticated physical ingredients move the predicted transparency always in one direction, making it larger.

There is a significant difference of behavior of transparency at lowest values of proton momentum or $Q^2$ between the results presented here and the ones of Ref.~\cite{Leitner:2009ke}, Fig. 3.
A maximum at $Q^2\sim 1 \ (\mathrm{GeV/c})^2$, which is seen there, comes from the bare model maximum at $p_p\sim 0.8 \ \mathrm{GeV/c}$, see Fig.~\ref{fig:trans_ingred}.
In our model, this structure mostly disappears when in-medium modification of the nucleon-nucleon cross sections are introduced.

\section{Discussion}
\label{sec:discussion}

The description of nuclear effects, and, in particular, multinucleon ejection mechanism, is one of the crucial uncertainties in the neutrino oscillation analyses.

Although, many theoretical~\cite{Martini:2010ex, Martini:2011wp, Nieves:2011pp, Megias:2016fjk, VanCuyck:2016fab, VanCuyck:2017wfn, Lovato:2017cux, Rocco:2018mwt, Mosel:2017zwq}, experimental~\cite{Abe:2016tmq, Abe:2017rfw, Patrick:2018gvi, Gran:2018fxa, Lu:2018stk}, and phenomenological~\cite{Gran:2013kda, Niewczas:2015iea, Lu:2015tcr} studies were done to increase the accuracy of the multinucleon ejection description, crude implementations in MC event generators limit the attempts of drawing conclusions using more exclusive final states, e.g., with one muon and proton in the final state.
In the context of NuWro, as the remaining models have either the satisfactory physical content (an exact SF implementation for the CCQE channel) or were successfully compared with data (for single pion production see Refs.~\cite{Gonzalez-Jimenez:2017fea, Nikolakopoulos:2018gtf}), one can attempt to investigate the seperation of multinucleon ejection events, assuming a proper control of the FSI modeling.
In the following subsections, we present two applications of the aforementioned cascade model uncertainties on the MC predictions in the CC$0\pi$ experimental channel.

\subsection{Application I: Single transverse variables}
\label{sec:sub:stv}

\begin{figure*}
 \includegraphics[width=\textwidth]{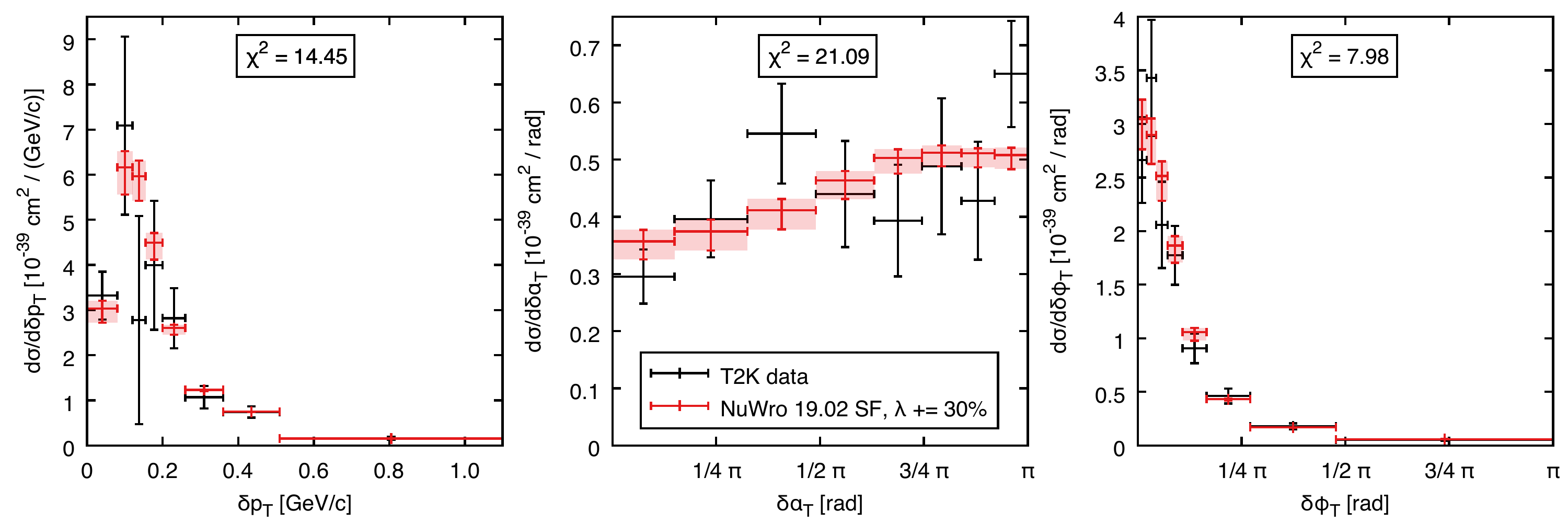}
 \caption{Impact of FSI uncertainty in NuWro predictions for single transverse variables. Experimental points are taken from~\cite{Abe:2018pwo}.}
 \label{fig:stv}
\end{figure*}

As the first application, we discuss T2K measurements of single transverse variables~\cite{Abe:2018pwo}.
These variables are defined in the following way
\begin{align}
    \delta p_\mathrm{T} & = |\delta \vec p_\mathrm{T}| = | (\vec{p}_p)_\mathrm{T} + \vec{k}^\prime_\mathrm{T} |, \\
    \delta \alpha_\mathrm{T} & = \arccos \frac{-\vec{k}^\prime_\mathrm{T} \cdot \delta \vec{p}_\mathrm{T}}{k^\prime_\mathrm{T} \delta p_\mathrm{T}}, \\
    \delta \phi_\mathrm{T} & = \arccos \frac{\vec{k}^\prime_\mathrm{T} \cdot (\vec{p}_p)_\mathrm{T}}{k^\prime_\mathrm{T} (p_p)_\mathrm{T}}.
\end{align}
Here $k^\prime$, $p_p$ correspond to the outgoing lepton and proton, and index $\mathrm{T}$ denotes the transverse projection w.r.t. the beam direction.
NuWro results are obtained with the SF model known to produce better results then LFG~\cite{Abe:2018pwo}.
Due to many adjustments in the FSI model, the results obtained with NuWro 19.02 differ notably from the ones of older versions of NuWro.
The most significant effect is an increase of normalization.
This does not change much the values of $\chi^2$ for the SF-based results, but makes the $\chi^2$ values larger for the LFG-based ones.

In Fig.~\ref{fig:stv}, we show how much of uncertainty comes from possible NuWro FSI mismodeling.
We see that applying a global decrease of the cascade mean free paths by 30\% decreases the normalization of the results.
We checked that this does not lead to a significant change of the calculated values of $\chi^2$.
Making mean free paths 30\% larger causes a slight increase of the value of $\chi^2$.
A general conclusion is that for single transverse variables, the error coming form FSI strength seems to be well under control.

\subsection{Application II: Proton multiplicities}
\label{sec:sub:multi}

\begin{figure}
 \includegraphics[width=0.5\textwidth]{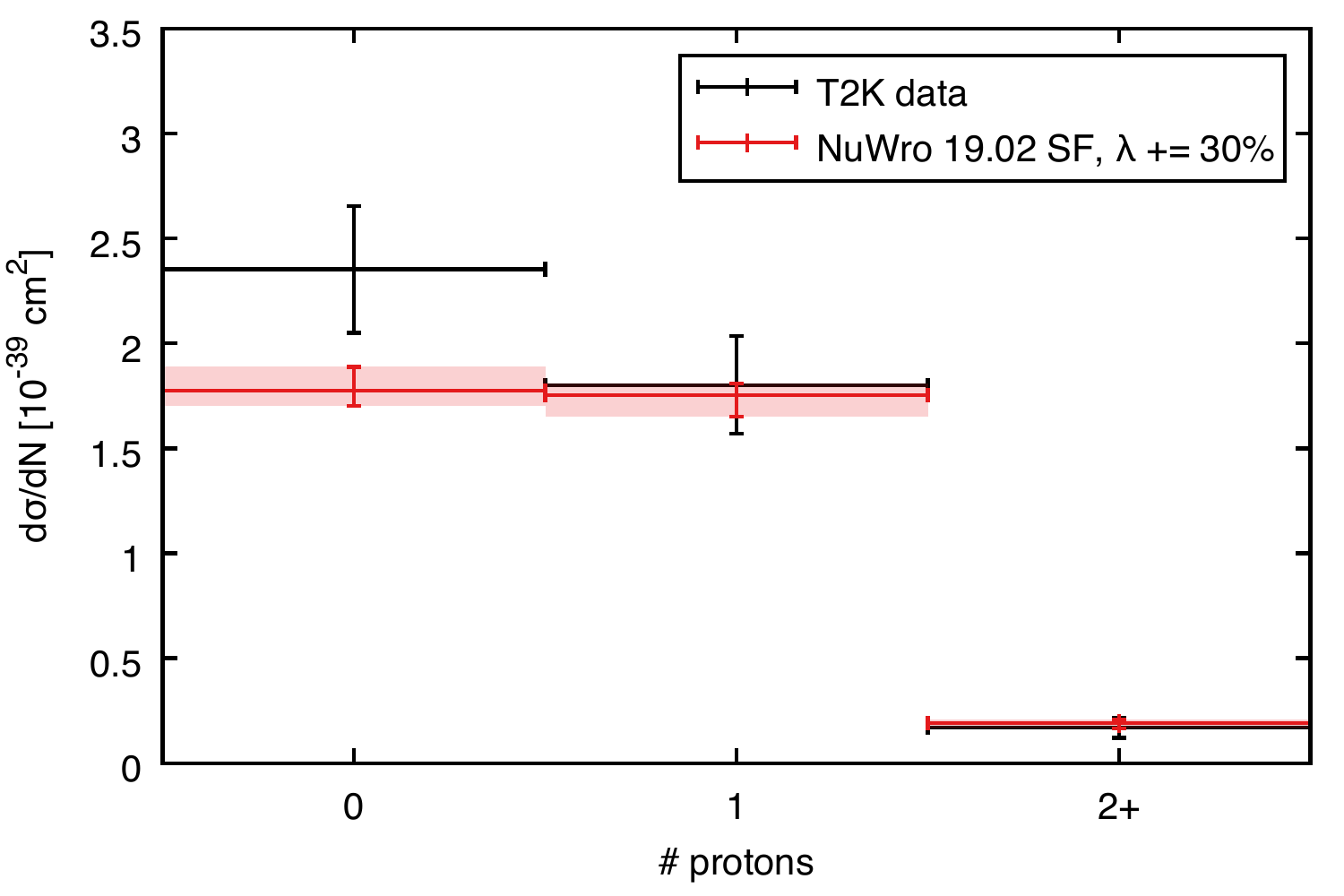}
 \caption{Impact of FSI uncertainty in NuWro predictions for proton multiplicities. Experimental points are taken from~\cite{Abe:2018pwo}.}
 \label{fig:multi}
\end{figure}

An observable that is potentially very sensitive to nucleon FSI effects is a distribution of number of reconstructed protons.
The dominant contribution to the experimental signal comes from CCQE events.
Thanks to FSIs, there is a fraction of CCQE events with more than one proton, otherwise such events would be impossible.
Another impact of FSI is that due to rescattering some protons loose kinetic energy dropping down below the detection threshold, what results in events with no detected protons.
In general, the FSI net effect is mostly a migration of events from N=1 to N=0.

In Fig.~\ref{fig:multi}, we show a comparison of NuWro predictions with the T2K data from Ref.~\cite{Abe:2018pwo}.
We see that the uncertainty coming from unknown strength of FSI is not large.
Here, larger nucleon mean free paths results in increasing proton multiplicities.
The data shape suggests that FSI strength should be set at the biggest value acceptable by the nuclear transparency data.
An impact of FSI on the distribution is smaller than expected.
It is because the experimental proton acceptance cuts eliminate most of events with FSI.

\section{Conclusions}
\label{sec:conclusions}

NuWro 19.02 features an improved nucleon cascade model that, using proper comparison tools, is able to reproduce nuclear transparency data, in particular in the energy region which is crucial in the context of \mbox{neutrino-nucleus} scattering physics.
The study presented in this paper shows that a cascade model should be enriched with many additional effects, such as, nucleon correlations, on top of a bare model with free \mbox{nucleon-nucleon} cross sections.

For the purpose of neutrino scattering physics, we estimated a $1\sigma$ error on the nucleon mean free paths in NuWro 19.02 with a result of 30\%.
This result was applied to recent T2K data that are potentially sensitive to nucleon FSI giving an uncertainty which suggests that FSI modeling in under control and that there should be other sources of data and MC disagreement that is still seen in NuWro results.
There is a solid foundation for using these datasets in future research of multinucleon ejection contributions and especially a very uncertain hadronic part of its modeling~\cite{Sobczyk:2012ms}.

All of the results obtained in this paper can be easily reproduced with any other neutrino MC generator, such as, NEUT or GENIE.
The outcome of this work should allow to further reduce systematic errors in the modeling of \mbox{neutrino-nucleus} scattering, and moreover, open a door for future analyses of more involved exclusive interaction channels.


\begin{acknowledgments}
We thank T.~Golan, C.~Juszczak, O.~Benhar, A.~Lovato, W.~Cosyn, N.~Jachowicz,~R.~Gonz\'alez~Jim\'enez, A.~Nikolakopoulos, J.~Zalipska and many our T2K Collaborators for suggestions and discussions at various stages of this work. We thank J. Pasternak for reading the manuscript of the paper. 
The authors were supported by NCN Opus Grant 2016/21/B/ST2/01092, and also by the Polish Ministry of Science and Higher Education, Grant DIR/WK/2017/05.
K.N. was partially supported by the Special Research Fund, Ghent University.
\end{acknowledgments}

\bibliography{refs}

\end{document}